# Hydrogen Storage on Platinum-Decorated Carbon Nanotubes with Boron, Nitrogen Dopants or Sidewall Vacancies

Jian-Ge Zhou and Quinton L. Williams

Department of Physics, Atmospheric Sciences and Geoscience, Jackson State University, Jackson, MS 39217, USA

**Abstract.** The interaction between hydrogen molecules and platinum (Pt)-decorated carbon nanotubes (CNTs) with boron (B)-, nitrogen (N)-dopants or sidewall vacancies is discussed from first-principle calculations. The adsorption patterns of hydrogen molecules on four types of Pt-decorated CNTs are investigated, and the partial density of states projected on the Pt atom is computed to reveal the response to the number of hydrogen molecules, dopants or vacancies. It is found that the B-, N-dopants or sidewall vacancies can adjust the binding energy between the hydrogen molecules and the Pt atom deposited on the defective CNT, while not reducing the maximum number of hydrogen molecules that are chemically adsorbed on the Pt atom. It is demonstrated that the binding energy of the first $H_2$ and the Pt atom on the pristine CNT or the CNT with the B-, N-dopants is quite strong, so each Pt atom in these three cases can only release the second $H_2$ under ambient conditions. However, when the Pt atom is deposited on the CNT with sidewall vacancies, it can adsorb and desorb two hydrogen molecules under ambient conditions.

**Introduction**

Unlike fossil fuels, hydrogen is a clean, environmentally friendly fuel, and expected to be widely used in the future energy infrastructure [1, 2]. Due to safety issues and relatively high energy costs, traditional storage approaches, such as compressed gas and liquefaction, are not suitable for $H_2$ [3]. Hydrogen as a commercial fuel requires that its storage can be processed more economically and conveniently than the state of the art. Carbon nanotubes (CNTs) are the promising candidate materials for the hydrogen storage because of their high surface to volume ratios, and high degree of reactivity between the carbon and hydrogen. Experiments indicated that the pristine CNTs can adsorb only about 1.0 wt % of $H_2$, and the binding energy is too small to meet the U.S. Department of Energy proposed goals which include that the binding energy per hydrogen molecule should be between 0.2 eV and 0.7 eV, and the $H_2$ storage capacity should exceed 6.0 wt % [4-7]. Experimentally, it was found that the lithium-, potassium- [8-10], nickel- [11], vanadium- [12] or palladium (Pd)-decorated [12-15] CNTs can enhance the hydrogen storage capacity. The Pd-decorated CNT can reach a high-capacity hydrogen storage of 8-12 wt % experimentally [15]. Density functional theory (DFT) simulations predicted that the platinum (Pt)-decorated CNT has higher hydrogen uptakes than the pristine CNT [16], and the titanium-decorated CNT can even adsorb up to 8.0 wt % of hydrogen [17]. DFT calculations further show that other metal decorated CNTs can increase the hydrogen storage capacity too [18-21]. However, on the metal decorated CNTs, the binding energy of the first $H_2$ is usually too strong to release the first hydrogen molecule under ambient conditions [16-21].

An alternative route is substitutional doping of CNTs with boron (B) or nitrogen (N) atoms, which modifies the electronic properties and reactivity of the pristine CNTs [22, 23]. The hydrogen storage on the B- or N-doped CNT has been studied experimentally [24, 25] and computationally [26-28]. The binding energy between the hydrogen molecules and the B- or N-doped CNT is still smaller than 0.2 eV, which indicates that the B- or N-doped CNT can not store hydrogen under ambient conditions. However, the B- or N-doped CNT manifests enhanced adsorption interaction with transition metal atoms [29-31], that is, the B and N dopants increase the stability of the transition metal deposition on the doped CNT [32]. The CNT with sidewall vacancies is more active than pristine one, and it has stronger interaction with transition metal atoms than the regular CNT [33]. A more stable adsorption of transition metal atoms on the CNT with B-, N-dopants or sidewall vacancies can reduce the loss of these transition metals, and their reduction might facilitate the commercialization of hydrogen as a more popular and cost-effective energy source.

Since a single free transition metal atom adsorbs more hydrogen molecules than that when it is deposited on a pristine CNT [16, 17], one may think that the increase of the adsorption energy between the transition metal atom and the defective CNT can reduce the capacity of the hydrogen storage. A moderate binding energy between the hydrogen molecules and the transition metal atom on the CNT is ideal for easy storage and release of $H_2$, but the binding energy of the first hydrogen molecule is too strong to release it under ambient conditions [16, 17]. Thus, we need to add some variables to the CNT so that we can adjust the binding energy to optimize the hydrogen uptakes. Recently, it was observed experimentally that the Pt atoms on the N-doped CNT enhance electron-transfer kinetics for the hydrogen evolution process [34], which sparks us to study the hydrogen storage on the transition metal decorated CNT with B-, N-dopants or sidewall vacancies. So far, what roles B-, N-dopants or sidewall vacancies in the transition metal decorated CNT play in the hydrogen storage capacity and the binding energy between $H_2$ and the transition metal atom is poorly understood, and it is imperative to answer the questions like: 1) Does the increase of the adsorption energy between the transition metal and the CNT with B-, N-dopants or sidewall vacancies reduce the capacity of the hydrogen storage? 2) Can dopants or vacancies adjust the binding energy of hydrogen molecules so that $H_2$ can adsorb and desorb under ambient conditions which requires the binding energy in the range of 0.2 eV - 0.7 eV per hydrogen molecule [35]?

To answer the above questions, in this article we choose a Pt-decorated CNT with B-, N-dopants or sidewall vacancies as a model to investigate the interaction between hydrogen molecules and the Pt atom from first-principle calculations. At first, we present adsorption energies and geometries for the Pt atom deposited on a (8,0) pristine CNT (**P**), B-doped CNT (**B**), N-doped CNT (**N**) and CNT with sidewall vacancies (**V**). It is demonstrated that the adsorption energies of the Pt atom on **B**, **N** and **V** types of CNTs are larger than that on the pristine CNT, which shows that the B-, N-dopants or sidewall vacancies of the CNT enhance the stability of the Pt deposition. Then we discuss the hydrogen molecule adsorption on the Pt atom deposited on **P**, **B**, **N** and **V** CNTs, respectively. We find that the first $H_2$ is partially dissociated and attached to the Pt atom on **P**, **B**, **N** and **V** CNTs. Later, we add the second hydrogen molecule to the system. Initially, we put the second $H_2$ near the CNT. After optimization, the second $H_2$ is partially dissociated with the help of the Pt atom. The adsorption of hydrogen molecules is mediated by weakening the bonds between the Pt atom and the CNT. Furthermore, we examine the addition of the third $H_2$ to the system, and find that the third $H_2$ is physisorbed to the Pt atom. To further explore the $H_2$ adsorption mechanism, we calculate the partial density of states (PDOS) projected on the Pt atom with zero, one and two $H_2$ on **P**, **B**, **N** and **V** CNTs to display the response to the number of hydrogen molecules, the B-, N-dopants or sidewall vacancies. We find that the increase of the adsorption interaction between the Pt and the CNT with the B-, N-dopants or sidewall vacancies does not change the maximum number of hydrogen molecules that are chemically adsorbed on the Pt atom. The binding energy of the first $H_2$ and the Pt atom is quite strong in the cases of **P**, **B**, and **N** CNTs, so each Pt atom can only release one $H_2$ under ambient conditions. However, when the Pt atom is deposited on **V** CNT, it can adsorb and desorb two hydrogen molecules under ambient conditions.

**Model System and Computational Method**

The calculations were carried out in the slab model by the unrestricted spin-polarized DFT [36, 37]. The electron-ion interaction has been described using the projector augmented wave (PAW) method

[38, 39]. The calculations have been performed by Perdew-Wang 91 (PW91) generalized gradient approximation (GGA) [40] and the local density approximation (LDA). The wave functions were expanded in a plane wave basis with an energy cutoff of 400 eV. Single-walled (8,0) CNT is taken as the model system. The supercell has 64 carbon atoms, and its size is $20 \times 20 \times 8.53$ Å with length of twice the periodicity of the (8,0) CNT. The CNT axis is set along z-direction. The lateral size of the CNT is 20 Å, which is large enough to minimize the interaction between adjacent tubes. The $k$ points were obtained from Monkhorst-Pack scheme, and $1 \times 1 \times 3$ $k$ point mesh was for the geometry optimization. The structures were fully optimized until the force on each atom was less than 0.02 eV/Å. Based on the optimized structures, the density of states was calculated using $1 \times 1 \times 20$ $k$-points. For the physisorption of hydrogen molecules, the $H_2$ binding energies obtained from LDA are in substantial agreement with experiments [41, 42]. The LDA has been preferred as a reliable and computationally efficient functional for systems involving the van der Waals interaction [43-46].

**Results and Discussion**

In this section, we first discuss the adsorption pattern of a Pt atom adsorbed on **P**, **B**, **N** and **V** CNTs. Then, we investigate the geometries and binding energies for one, two and three hydrogen molecules attached to the Pt atom on **P**, **B**, **N** and **V** types of the CNTs, respectively. Various orientations of the Pt atom and the hydrogen molecules, as well as different adsorption sites, have been exploited to find out the lowest energy configuration for each case. Thus the obtained structure can be regarded as the global energy minimum through broad search for local minima on the potential energy surface of the adsorption interaction. Finally, we compute the PDOS projected on the Pt atom to explain different adsorption patterns. The units for the bond lengths and binding energies listed in the following tables are Angstrom (Å) and eV.

**Pt on P, B, N and V CNTs.** First, let us begin with our analysis with the geometries and adsorption energies of the optimized structures for the Pt atom on **P**, **B**, **N** and **V** CNTs, as displayed in Table 1.

Table 1: The geometries and adsorption energies for the structures of a Pt atom adsorbed on a (8,0) pristine CNT (**P**), B-doped CNT (**B**), N-doped CNT (**N**) and a CNT with a vacancy (**V**), respectively. The entries $d_{Pt-C1}$, $d_{Pt-C2}$, $d_{Pt-C3}$, $d_{Pt-B}$ and $E_{ads}^{Pt}$ refer to the bond lengths of Pt-C1, Pt-C2, Pt-C3, Pt-B and the adsorption energy between the Pt atom and the substrate.

| bond length & adsorption energy | P CNT | B CNT | N CNT | V CNT |
|---|---|---|---|---|
| $d_{Pt-C1}$ | 2.05 | - | 2.07 | 1.98 |
| $d_{Pt-C2}$ | 2.05 | 2.04 | 2.02 | 1.90 |
| $d_{Pt-C3}$ | - | - | - | 1.98 |
| $d_{Pt-B}$ | - | 2.15 | - | - |
| $E_{ads}^{Pt}$ | 2.49 | 2.85 | 2.76 | 5.99 |
| $\Delta E_{ads}^{Pt}$ ( $E_{ads}^{Pt}$ - $E_{prinstine}^{Pt}$) | 0 | 0.36 | 0.27 | 3.50 |

To obtain the most stable structure in each case, all the possible symmetric sites (including the bridge site over axial C-C bond (**BA**), the bridge site over a zigzag C-C bond (**BZ**) (Fig. 1a), the hollow and atop sites) have been considered as the initial positions of B, N and Pt atom. The adsorption energy of the Pt atom on a CNT is defined as $E_{ads}^{Pt} = E_{Pt} + E_{a\ given\ CNT} - E_{a\ given\ CNT\ +\ Pt}$. The "a given CNT" stands for one of **P**, **B**, **N** and **V** CNTs. The entries $d_{Pt-C1}$, $d_{Pt-C2}$, $d_{Pt-C3}$, $d_{Pt-B}$ and $E_{ads}^{Pt}$ refer to the bond lengths of Pt-C1, Pt-C2, Pt-C3, Pt-B, and the adsorption energy between the Pt atom and the substrate, where C1, C2 and C3 are denoted in Fig. 1. In an N-doped CNT, the Pt atom forms bonds with two carbons nearby the doped nitrogen (Fig. 1c), and does not form a direct bond with N atom, so in the Tables, we do not include $d_{Pt-N}$.

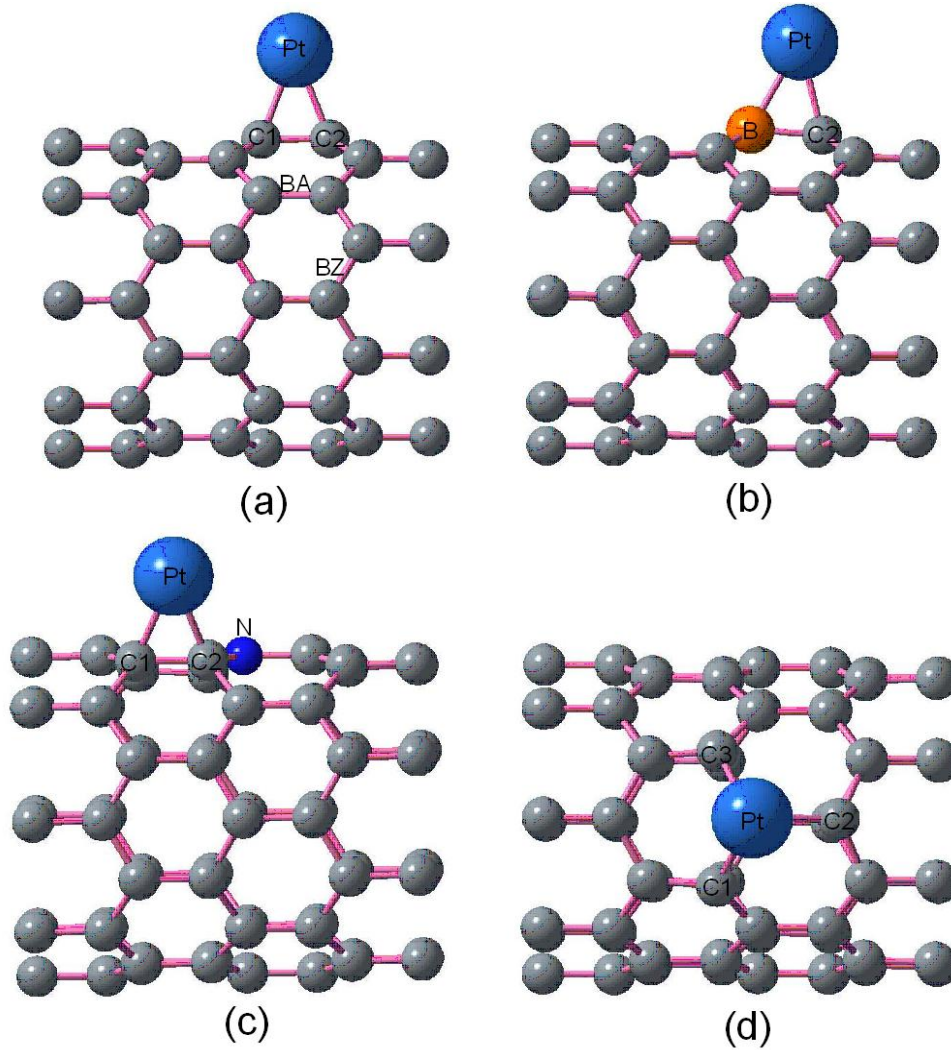

Fig. 1: (a) Pt atom adsorbed on **P** CNT. (b) **B** CNT. (c) **N** CNT. (d) **V** CNT.

The adsorption energies in Table 1 show that the interaction of the adsorbed Pt atom on **B**, **N** or **V** CNT is stronger than that on the pristine CNT, with energy increases of +0.36eV (**B**), +0.27eV (**N**) and +3.50eV (**V**). The B-, N-dopants and sidewall vacancies of the CNT increase the relative reactivity with respect to the pristine CNT. A Pt atom sits on **BA** instead of **BZ** in the (8, 0) pristine CNT (Fig. 1a). On the **B** type CNT, the Pt is located between the boron and carbon, i.e., **BA** site (Fig. 1b), so the enhancement in Pt adsorption on the **B** CNT is attributed to the strong hybridization between the Pt d-orbitals and boron p-orbitals. For an N-doping in the CNT, the Pt atom is on the **BA** site nearby the nitrogen, which can be interpreted as that the nitrogen atom activates the nitrogen-neighboring carbon atoms and mediates the enhancement of the Pt adsorption (Fig. 1c). In the case of the sidewall vacancies, the Pt atom sits above each vacancy and forms hybrid bonds with three carbons (Fig. 1d), which is the strongest adsorption among four kinds of the CNTs. The bond length between Pt and C is around 1.95Å (**V**), and smaller than those of the Pt atom on **P** (2.05Å), **B** (2.04Å) and **N** CNTs (2.02Å).

**One $H_2$ adsorbed to Pt on P, B, N and V CNTs.** Next, we investigate how the first hydrogen molecule interacts with the Pt atom deposited on **P**, **B**, **N** and **V** CNTs, respectively. The binding energy for the first hydrogen molecule is given by $E_{bind}^{(1)} = E_{H_2} + E_{a\ given\ CNT\ +\ Pt} - E_{a\ given\ CNT\ +\ Pt\ +\ 2H}$. Initially, we placed a hydrogen molecule near one of the Pt-decorated **P**, **B**, **N** and **V** CNTs. After optimization, the $H_2$ is chemisorbed to the Pt atom on **P**, **B**, **N** and **V** types of the CNTs without crossing any energy barrier. The corresponding optimized configurations are described in Fig. 2.

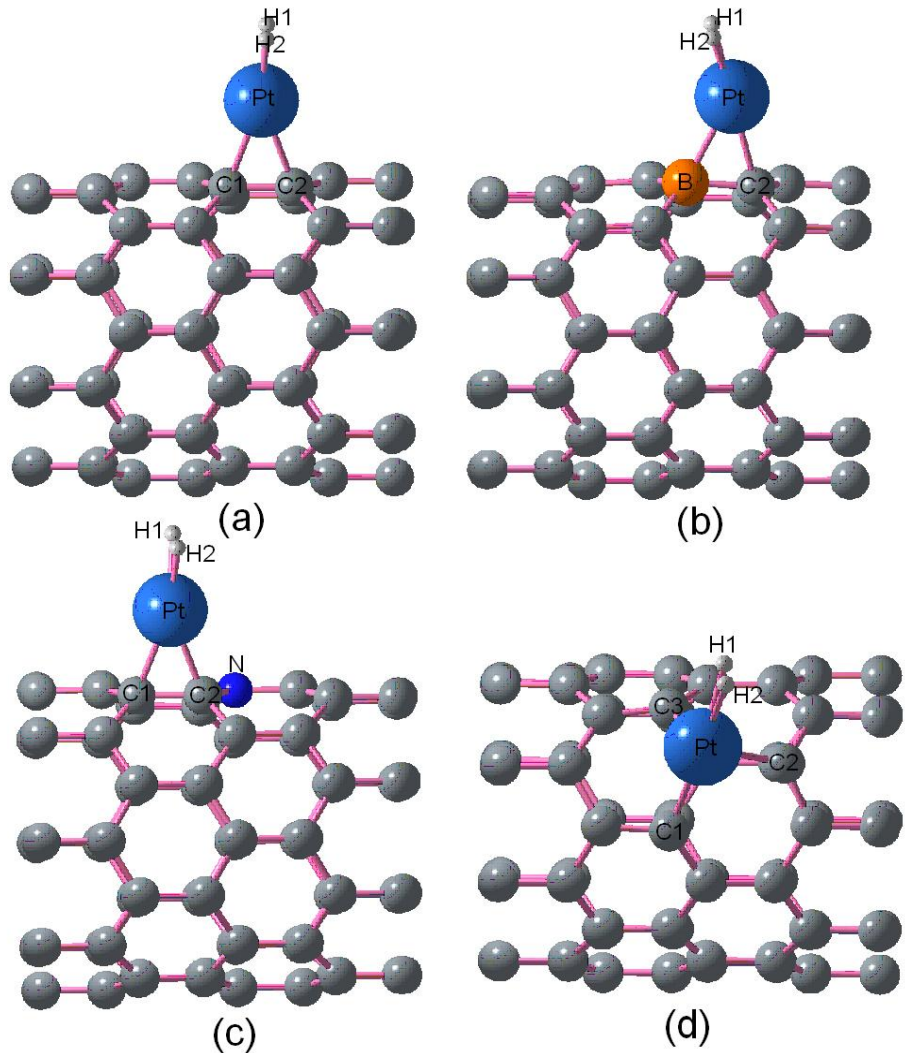

Fig. 2: (a) One $H_2$ attached to the Pt atom on **P** CNT. (b) **B** CNT. (c) **N** CNT. (d) **V** CNT.

The geometries and binding energies for the first $H_2$ attached to the Pt atom on **P**, **B**, **N** and **V** types of the CNTs are listed in Table 2. Since the H-H bond length of the isolated $H_2$ is 0.75Å, the partial dissociation of the hydrogen molecule is displayed via the elongated H-H bond lengths: 0.98Å (**P**), 1.02Å (**B**), 0.96Å (**N**) and 0.87Å (**V**), which are consistent with their binding energies. The partially dissociated hydrogen molecule interacts with the Pt atom via the Kubas interaction which is the hybridization of the d states of the Pt atom and the s state of the $H_2$ [47]. There are locally stable structures in which the adsorbed hydrogen is in the atomic form, however, their binding energies are less than the above ones where the $H_2$ is partially dissociated (Fig. 2). To compare the interaction between the Pt atom and the CNTs for the cases without and with a hydrogen molecule, we display the bond lengths of Pt-C2: (2.05Å, 2.15Å, **P**), (2.04Å, 2.14Å, **B**), (2.02Å, 2.14Å, **N**) and (1.90Å, 1.93Å, **V**), where the first number in the parentheses stands for the Pt-C2 bond length without the hydrogen molecule, the second corresponds to the Pt-C2 bond length with one $H_2$, and the letters represent four types of the CNTs. By comparison, we find that the adsorption of the $H_2$ weakens the bonds between the Pt and the CNT. The Pt atom forms the strongest bond with **V** CNT ($E_{ads}^{Pt}$=5.99eV), thus the corresponding bond between the Pt and the $H_2$ is the weakest one in four types of the CNTs ($E_{bind}^{(1)}$=0.38eV, **V**), that is, the first $H_2$ is molecularly chemisorbed to the Pt atom on **V** CNT. Comparing the binding energies of the first $H_2$ on the Pt atom on four types of the CNTs, we find that the first H$_2$ can be adsorbed and desorbed easily on the Pt atom on **V** CNT.

Table 2: The geometries and binding energies for one H$_2$ attached to the Pt atom on **P**, **B**, **N** and **V** CNTs.

| bond length & binding energy | **P** CNT | **B** CNT | **N** CNT | **V** CNT |
|---|---|---|---|---|
| $d_{Pt-C1}$ | 2.16 | - | 2.14 | 1.99 |
| $d_{Pt-C2}$ | 2.15 | 2.14 | 2.14 | 1.93 |
| $d_{Pt-C3}$ | - | - | - | 2.05 |
| $d_{Pt-B}$ | - | 2.21 | - | - |
| $d_{H1-Pt}$ | 1.66 | 1.64 | 1.67 | 1.82 |
| $d_{H2-Pt}$ | 1.66 | 1.65 | 1.67 | 1.81 |
| $d_{H1-H2}$ | 0.98 | 1.01 | 0.96 | 0.87 |
| $E_{bind}^{(1)}$ | 1.36 | 1.33 | 1.22 | 0.38 |

**Two and three H$_2$ adsorbed to Pt on P, B, N and V CNTs.** Let us add the second hydrogen molecule to the system. Initially, we put the second H$_2$ near the CNT. The optimized structures for two hydrogen molecules attached to the Pt atom on **P**, **B**, **N** and **V** CNTs are illustrated in Fig. 3. The bond lengths and binding energies for the system with two H$_2$ attached to the Pt atom on **P**, **B**, **N** and **V** types of the CNTs are displayed in Table 3, where the binding energy for the second hydrogen molecule is defined as $E_{bind}^{(2)} = E_{H_2} + E_{a\,given\,CNT\,+\,Pt\,+\,2H} - E_{a\,given\,CNT\,+\,Pt\,+\,2H\,+\,2H}$. After optimization, the bond length of the second hydrogen molecule (H3-H4) becomes 0.90Å (**P**), 0.86Å (**B**), 0.84Å (**N**) and 0.85Å (**V**); that is, the second H$_2$ is partially dissociated without crossing any energy barrier. The partial dissociation of the second H$_2$ is mediated by further weakening the bonds between the Pt atom and the CNT. This further elongation of Pt-C2 is revealed through the bond length increase: (2.15Å, 2.16Å, **P**), (2.14Å, 2.31Å, **B**) and (2.14Å, 2.28Å, **N**), (1.93Å, 1.93Å, **V**), where the first number in the parentheses is for the Pt-C2 bond length with one hydrogen molecule, and the second number represents the Pt-C2 bond length with two H$_2$. Here, we point out that in the case of **V** CNT, the Pt-C2 bond length does not change, but the bonds Pt-C1 and Pt-C3 elongate.

The molecular chemisorption of the second hydrogen molecule to the Pt atom can affect the magnitude of the dissociation of the first H$_2$. For **P** CNT, the bond length of the first hydrogen molecule (H1-H2) decreases from 0.98Å (one H$_2$) to 0.89Å (two H$_2$). The H1-H2 bond length only changes slightly for **V** CNT, from 0.87Å to 0.85Å. However, they change drastically in **B** and **N** CNTs (1.01Å →1.94Å, 0.96Å →2.05Å), so the second hydrogen molecule makes the first one further dissociated. The binding energies indicate that **B** CNT has the highest binding energy (0.63eV) for the second H$_2$ among four types of the CNTs. The B-, N-dopants or sidewall vacancies of the CNT can increase the interaction between the Pt atom and the second H$_2$. After adsorption, the four structures of two hydrogen atoms on the Pt atom on **P**, **B**, **N** and **V** CNTs are very different (Fig. 3), which shows that the dopants and vacancies modify the hydrogen adsorption pattern.

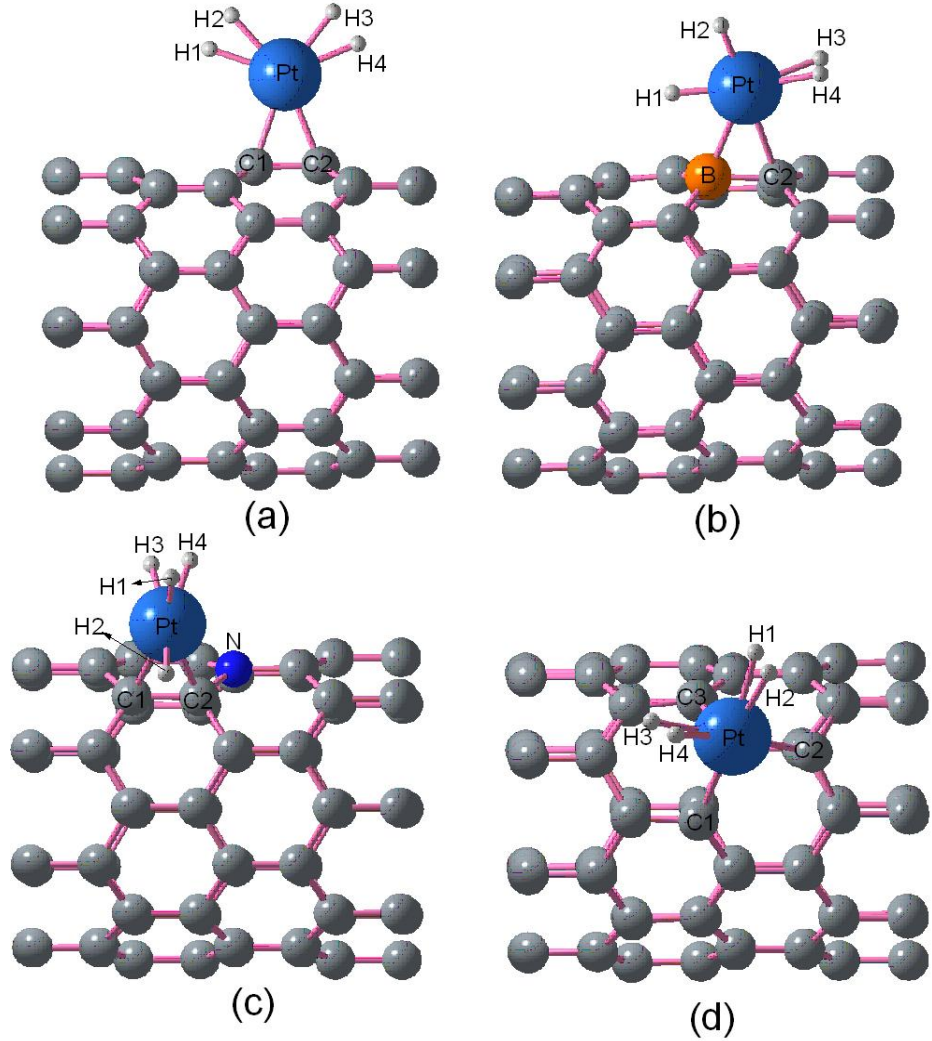

Fig. 3: (a) Two H$_2$ attached to the Pt atom on **P** CNT. (b) **B** CNT. (c) **N** CNT. (d) **V** CNT.

Table 3: The geometries and binding energies for two H$_2$ attached to the Pt atom on **P**, **B**, **N** and **V** CNTs.

| bond length & binding energy | P CNT | B CNT | N CNT | V CNT |
|---|---|---|---|---|
| $d_{Pt-C1}$ | 2.15 | - | 2.32 | 2.02 |
| $d_{Pt-C2}$ | 2.16 | 2.31 | 2.28 | 1.93 |
| $d_{Pt-C3}$ | - | - | - | 2.10 |
| $d_{Pt-B}$ | - | 2.28 | - | - |
| $d_{H1-Pt}$ | 1.76 | 1.58 | 1.59 | 1.84 |
| $d_{H2-Pt}$ | 1.76 | 1.59 | 1.58 | 1.84 |
| $d_{H3-Pt}$ | 1.75 | 1.77 | 1.81 | 1.88 |
| $d_{H4-Pt}$ | 1.75 | 1.77 | 1.90 | 1.84 |
| $d_{H1-H2}$ | 0.89 | 1.94 | 2.05 | 0.85 |
| $d_{H3-H4}$ | 0.90 | 0.86 | 0.84 | 0.85 |
| $E_{bind}^{(2)}$ | 0.33 | 0.63 | 0.60 | 0.47 |

If we add the third hydrogen molecule to the above system, we find that the third H$_2$ is only physisorbed to the Pt atom. The optimized structures are depicted in Fig. 4. To evaluate the van der Waals correction, we apply LDA to optimize the structures with the third H$_2$ attached to the Pt atom on **P**, **B**, **N** and **V** CNTs [41, 42]. The distance between the center of the third H$_2$ and the Pt atom is

2.67Å (**P**), 2.60Å (**B**), 2.58Å (**N**) and 2.51Å (**V**). The corresponding binding energy $E_{bind}^{(3)}$ is 0.04 eV (**P**), 0.05 eV (**B**), 0.04 eV (**N**) and 0.03 eV (**V**). These results indicate that up to two hydrogen molecules can be chemically adsorbed on a Pt atom and the third $H_2$ is attached to the Pt atom by weak physisorption bonds. If we placed the forth $H_2$ to the system, after optimization it escaped away from the Pt atom.

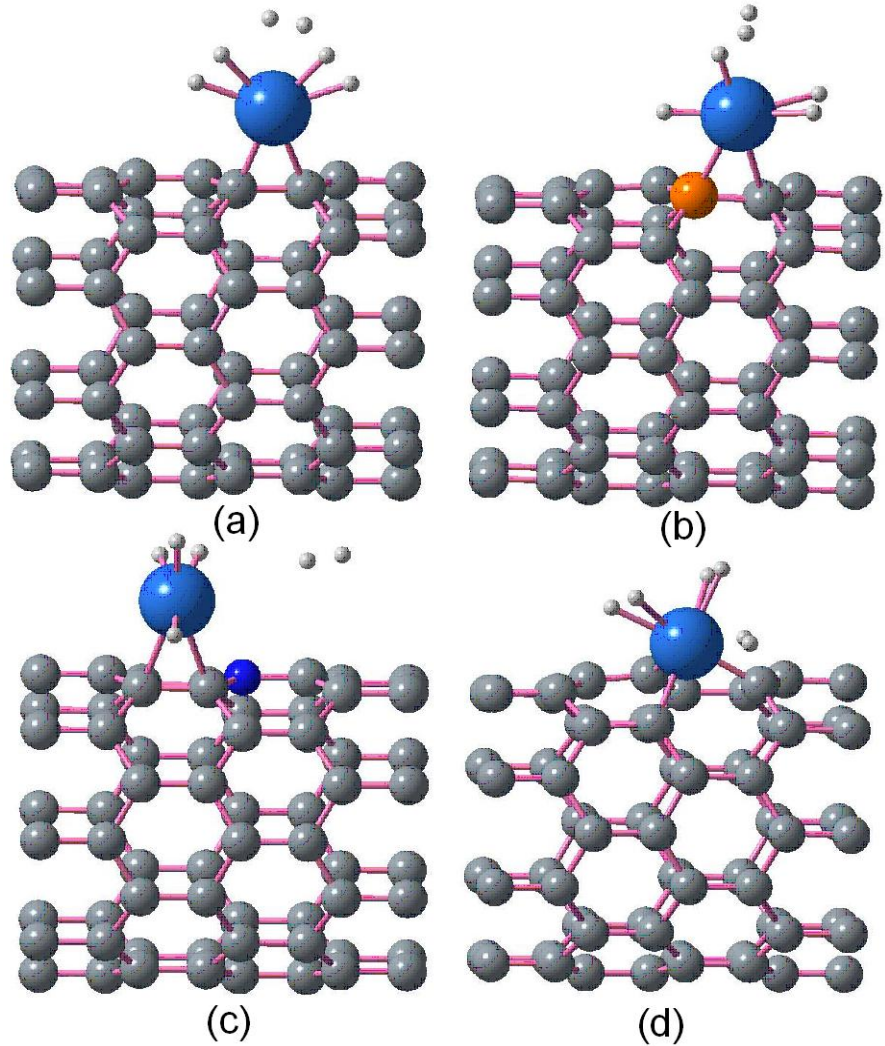

Fig. 4: (a) Three $H_2$ attached to the Pt atom on **P** CNT. (b) **B** CNT. (c) **N** CNT. (d) **V** CNT.

The binding energies of the third $H_2$ are much less than 0.2 eV, so it is weak to adsorb the third $H_2$ to the Pt atom [20, 35]. The binding energies of the first and second $H_2$ for four types of the CNTs are: (1.36 eV, 0.33 eV, **P**), (1.33 eV, 0.63 eV, **B**), (1.22 eV, 0.60 eV, **N**) and (0.38 eV, 0.47 eV, **V**), where the first number in the parentheses represents the binding energy of the first $H_2$ ($E_{bind}^{(1)}$) and the second number stands for the binding energy of the second $H_2$ ($E_{bind}^{(2)}$) (see Tables 2 and 3). These binding energies show that in the cases of **P**, **B** and **N** CNTs, each Pt atom can only adsorb and desorb the second $H_2$ under ambient condition which requires the binding energies in the range of 0.2 eV - 0.7 eV per hydrogen molecule [20, 35]. Since the binding energies of the first and second $H_2$ attached to the Pt atom on **V** CNT are 0.38 eV and 0.47 eV, each Pt atom on **V** CNT can adsorb and desorb two hydrogen molecules under ambient conditions, thus the sidewall vacancies of the CNT make a Pt atom to store and release hydrogen molecules more efficiently.

**PDOS projected on the Pt atom.** To elucidate how the B-, N-dopants, sidewall vacancies and the number of hydrogen molecules play roles in the adsorption process, we calculate the PDOS projected on the Pt atom on **P**, **B**, **N** and **V** CNTs (Fig. 5).

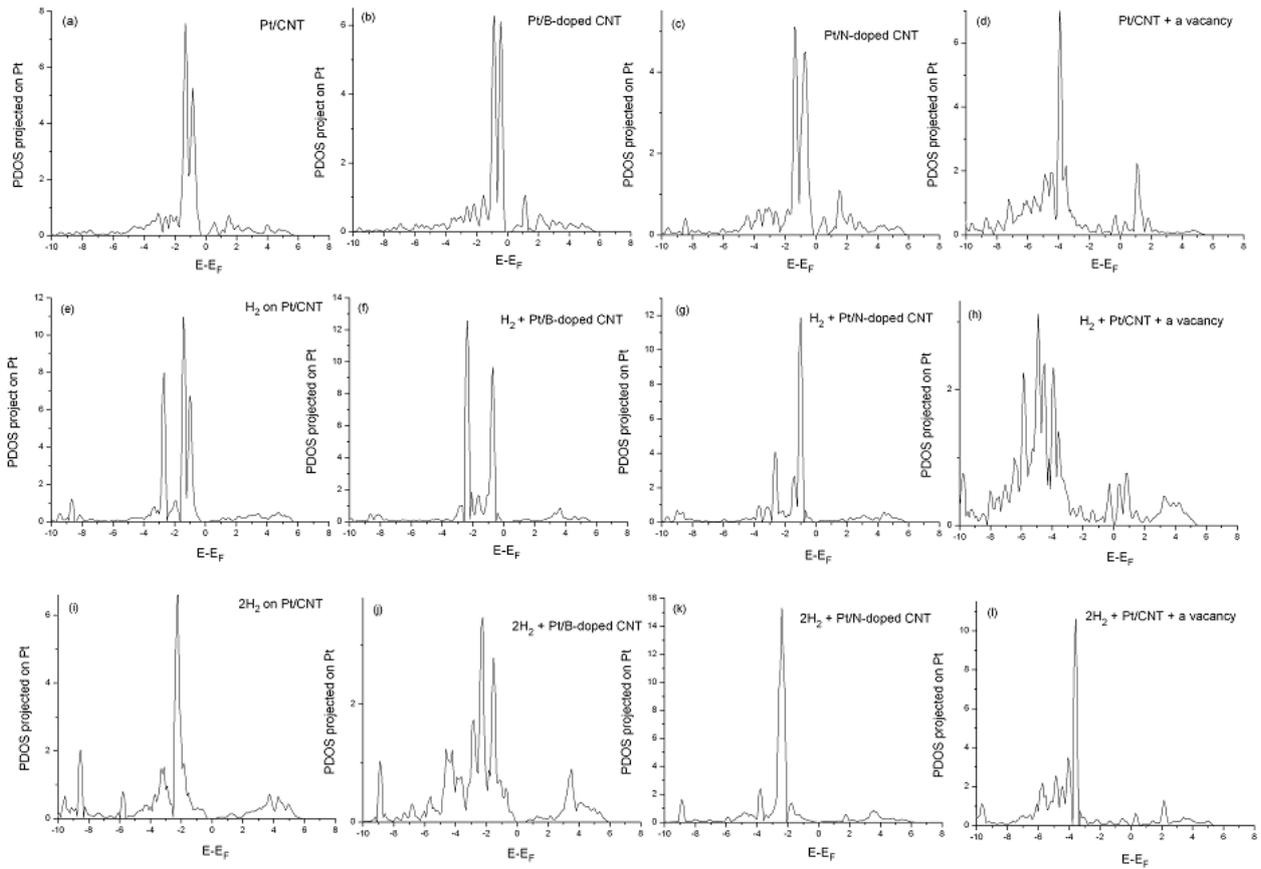

Fig. 5: (a) PDOS projected on the Pt atom adsorbed on **P** CNT. (b) **B** CNT. (c) **N** CNT. (d) **V** CNT. (e) PDOS projected on the Pt atom adsorbed to **P** CNT and one $H_2$. (f) **B** CNT and one $H_2$. (g) **N** CNT and one $H_2$. (h) **V** CNT and one $H_2$. (i) PDOS projected on the Pt atom adsorbed to **P** CNT and two $H_2$. (j) **B** CNT and two $H_2$. (k) **N** CNT and two $H_2$. (l) **V** CNT and two $H_2$.

The stronger adsorption energies of the Pt atom on the CNTs usually correspond to larger overlap between the Pt d-states and the p-states of the CNTs below the Fermi level. By comparing the primary peaks around -1 eV in **B** and **N** CNTs with those in **P** CNT (Fig. 5a, Fig. 5b and Fig. 5c), we find that the hybridization of the Pt d-states with p-states of **B** or **N** CNT is stronger than with p-states of **P** CNT, that is, the B- and N-dopants enhance the hybridization. The primary peaks of the Pt PDOS of **V** CNT turn out to be at lower energy than those of **P**, **B** and **N** CNTs, and therefore the Pt atom forms the strongest bond with **V** CNT among four types of CNTs (Fig. 5d). If one $H_2$ is added, the Pt d-states on **P** CNT interacts with the p-states of **P** CNT (peaks around -1 eV) and s-states of $H_2$ (peaks around -9 eV) (Fig. 5e), that is, the peaks around -9 eV correspond to the hybridization of the Pt d-states with the hydrogen s-states. Comparing the peaks around -9 eV, we see that the interaction between the Pt atom on **P** CNT and the $H_2$ is the strongest and the one between the Pt on **V** CNT and the $H_2$ is the weakest, which is in good agreement with calculated binding energies (Fig. 5e - Fig. 5h). When two $H_2$ are adsorbed, the Pt is almost saturated, i.e., the s- and d-orbitals of the Pt atom are nearly completed. The altitudes of -10 eV peak for one and two $H_2$ attached to the Pt on **V** CNT are 0.7 and 1.1, which are comparable with $E_{bind}^{(1)} = 0.38$ eV and $E_{bind}^{(2)} = 0.47$ eV (Fig. 5h, Fig. 5l). Because the interaction between the third $H_2$ and the Pt atom is the weak van der Waals force, the Pt PDOS with three $H_2$ is similar to that with two $H_2$.

**Conclusion**

We have examined the interaction between hydrogen molecules and the Pt atom on the CNT with the

B-, N-dopants or sidewall vacancies from first-principle calculations. It has been demonstrated that the adsorption energies of the Pt atom on **B**, **N** and **V** types of CNTs are larger than that on the pristine CNT, which shows that the B-, N-dopants or sidewall vacancies of the CNT can enhance the stability of the Pt deposition. Our computational results for the hydrogen molecule adsorption on the Pt atom deposited on **P**, **B**, **N** and **V** CNTs have shown that the first $H_2$ is partially dissociated and attached to the Pt atom on **P**, **B**, **N** and **V** CNTs. When we put the second $H_2$ near the CNT, after optimization, we have revealed that the second $H_2$ is partially dissociated with the help of the Pt atom. The dissociation of hydrogen molecules is mediated by weakening the bonds between the Pt atom and the CNT. Furthermore, we have investigated the addition of the third $H_2$ to the system, and found that the third $H_2$ is only weakly attached to the Pt atom by van der Waals force. To elucidate the $H_2$ adsorption mechanism, we have calculated the PDOS projected on the Pt atom with zero, one and two $H_2$ on **P**, **B**, **N** and **V** CNTs to display the response to the number of hydrogen molecules, dopants or vacancies. The results have displayed that the dopants and vacancies of the CNTs results in different hydrogen adsorption structures on the Pt atom on **P**, **B**, **N** and **V** CNTs. We have also discovered that the increase of the adsorption interaction between the Pt and the CNT with the B-, N-dopants or sidewall vacancies does not change the maximum number of hydrogen molecules that are chemically adsorbed on the Pt atom. In other words, the dopants and vacancies can adjust the binding energy between the hydrogen molecules and the Pt atom deposited on the defective CNT, while not reducing the capacity of the hydrogen storage. In the cases of **P**, **B** and **N** CNTs, the binding energy of the first $H_2$ with the Pt atom is too strong to release the first $H_2$, so each Pt atom can only release the second $H_2$ under ambient conditions. However, when the Pt atom is deposited on **V** CNT, it can adsorb and desorb two hydrogen molecules easily, thus from the binding energy point of view, the sidewall vacancies can enhance the capacity of the hydrogen storage under ambient conditions. In the above, we only considered a single B, N or Pt atom, issues like the clustering of B, N or Pt atoms on the CNT and how the aggregation affects the hydrogen storage will be discussed in near future.

## Acknowledgments

This work is funded in part by the DoD (Grant No. W912HZ-06-C-0057).